# Crystallographic-orientation dependence of the early-stage oxidation of Zr single crystals: an XPS study of suboxide formation and in-depth distribution

J. Sacanell[a,b,*] and F. Conde[a]

[a] *Departamento de Física de la Materia Condensada, Centro Atómico Constituyentes, Comisión Nacional de Energía Atómica (CNEA), Av. General Paz 1499, (1650) San Martín, Buenos Aires, Argentina*

[b] *Instituto de Nanociencia y Nanotecnología, CNEA-CONICET, Argentina*

* *Corresponding author. E-mail: joaquinsacanell@cnea.gob.ar*

## Abstract

The influence of crystallographic orientation on the early stages of the oxidation of pure Zr single crystals was studied by X-ray photoelectron spectroscopy (XPS). Two samples, cut from the same α-Zr single crystal, were oxidized simultaneously at room temperature and $O_2$ pressures of $1 \times 10^{-8}$, $1 \times 10^{-7}$ and $1 \times 10^{-6}$ torr: one with its surface normal parallel to the c-axis (Z1, basal (0001) plane) and the other with its normal 12° from that of a prismatic plane (Z2, near-prismatic orientation). At all three pressures the oxidation kinetics of both samples followed a three-stage, logarithmic-type behaviour, but the near-prismatic sample (Z2) incorporated oxygen faster and to a greater extent than the basal sample (Z1) in every case. Deconvolution of the Zr 3d core-level spectra resolved, in addition to metallic Zr and $ZrO_2$, two sub-stoichiometric Zr–O compounds, denoted (ZrO)a and (ZrO)b, with binding-energy shifts of 1.3 and 2.3 eV with respect to metallic Zr, consistent with sub-oxides previously reported for polycrystalline Zr. Angle-resolved XPS showed that $ZrO_2$ is the outermost compound in both orientations, while the two sub-oxides are distributed nearly homogeneously through the film. Oxide thicknesses, calculated from the attenuation of the metallic Zr 3d signal, ranged from 10 to 13 Å for Z1 and from 14 to 19 Å for Z2 depending on the oxidation pressure, in close agreement with earlier measurements on cold-worked polycrystalline Zr, whose surface texture is dominated by prismatic-oriented grains.

## 1. Introduction

Zirconium and several of its alloys are used as fuel-cladding materials in power reactors because of their low thermal-neutron absorption cross-section and their good corrosion resistance under reactor operating conditions. In service, the cladding surface is protected by a thin, dense oxide film that forms spontaneously on contact with the coolant and acts as a barrier against further ingress of oxidizing species. Understanding how this film nucleates and grows from the earliest stages of exposure to oxygen, well before it reaches a thickness accessible to bulk techniques, has therefore been of long-standing interest to the nuclear-materials community [1–6].

Surface-sensitive electron spectroscopies, in particular X-ray photoelectron spectroscopy (XPS) and Auger electron spectroscopy (AES), made it possible to follow the first stages of the Zr–O interaction directly, revealing not only the growth kinetics of the oxide but also the presence of Zr–O compounds with an oxidation state intermediate between metallic Zr and stoichiometric $ZrO_2$ [2,4,5,7,8,9]. The chemical nature and the exact number of these sub-oxides, however, has never been fully settled: some authors report a single intermediate ZrOx compound whose apparent binding energy must be allowed to vary with exposure to fit the data [2], while others resolve two [4,5] or three [3,10] discrete sub-oxide components. Most of this body of work was carried out on polycrystalline Zr, and only a handful of studies addressed single-crystal surfaces, generally focusing on one orientation at a time [1,6]. An early study by Pemsler already reported an effect of crystallographic orientation on the oxidation kinetics of Zr [11], although without a direct, side-by-side comparison of two independently indexed single-crystal orientations under identical conditions. Also, a comparison of Zr(0001) and Zr(10-10) oxidation kinetics was performed by Bakradze et al. [12], mainly focused on real-time spectroscopic ellipsometry over a range of elevated temperatures (300–450 K) at a single, fixed oxygen dose.

This is a relevant gap because the surfaces of cold-worked Zr cladding tubes are strongly textured: rolling and annealing produce a majority of grains with their basal poles tilted away from the tube normal and their prismatic planes preferentially exposed at the surface. If oxidation proceeds at different rates on the basal and prismatic planes of the hexagonal-close-packed α-Zr structure, the macroscopic corrosion behaviour of a component should depend on its crystallographic texture. Anisotropic, texture-dependent oxide formation on differently oriented grains of polycrystalline Zr has long been suspected and continues to be discussed in recent reviews of Zr-alloy corrosion [13]. However, a room-temperature, exposure-resolved comparison of the sub-oxide composition evolving on two differently oriented single crystals, including an orientation representative of the near-prismatic texture typical of cold-worked cladding tubes, was still lacking.

The work reported here addresses this question using XPS to compare, under strictly identical conditions, the room-temperature oxidation of two α-Zr single crystals cut from the same parent crystal: one oriented with the basal (0001) plane at the surface (sample Z1) and the other close to a prismatic plane (sample Z2). Both samples were oxidized simultaneously at three $O_2$ pressures spanning two orders of magnitude ($1 \times 10^{-8}$ to $1 \times 10^{-6}$ torr), and the oxidation kinetics, the sub-oxide composition of the growing film, its in-depth distribution (via angle-resolved XPS) and its final thickness were determined and compared between orientations.

The question addressed here has not been resolved by the decades of work that followed. A density-functional-theory study of the anisotropy of Zr oxidation across different low-index surfaces [14] confirmed, from a theoretical standpoint, that facet-dependent oxidation is to be expected on α-Zr, while atomic-scale experimental studies combining atom-probe tomography, aberration-corrected electron microscopy and multimodal chemical imaging have continued to probe the earliest, sub-nanometre stages of Zr-alloy oxidation and the atomic structure of the metal–oxide interface [15,16] without settling the question of orientation dependence. The chemical identity of the Zr sub-oxide phases remains an active topic in its own right: as recently as 2022, two independent studies debated the presence and structural role of a ZrO sub-oxide phase at the metal–oxide interface of corroded Zr alloys [10,17], and long-term oxidation studies under simulated reactor-coolant conditions continue to appear in the literature [18,19]. In a 2026 review of Zr-alloy oxidation for nuclear fuel cladding, the identity and mechanistic role of the sub-oxide phases at the metal–oxide interface, and the extent to which crystallographic texture controls the oxidation rate, are both still described as unresolved and, in places, contradictory across the literature [20]. The present results, obtained on a directly paired set of well-characterised single-crystal orientations, are therefore still relevant to that open discussion.

## 2. Experimental

### 2.1. Samples

α-Zr single crystals were grown from a 12 mm diameter, 4 mm thick rod of 99.99% pure Zr (main impurities, in ppm: Al 5, S 1, K 1.5, Fe 2, Cu 2.2, Si 5, Sn 2) by thermal cycling around the α↔β transformation temperature (862 °C), a method known to produce untensioned single, bi- or tri-crystals of α-Zr [21]. The crystallographic orientation of the resulting grains was determined by back-reflection Laue X-ray diffraction. Two samples were then cut from the same crystal by electro-discharge machining: sample Z1, with its surface normal parallel to the c-axis of the hexagonal-close-packed lattice (i.e. exposing the basal (0001) plane), and sample Z2, with its surface normal 12° from the normal of a prismatic plane, as established from its own Laue pattern (i.e. close to, but not exactly on, a prismatic plane), which is the basis for the basal/near-prismatic comparison presented in this work. The single crystals Z1 and Z2 were obtained from the same stock of α-Zr single crystals grown by thermal cycling through the α↔β transformation temperature at this institution; the growth procedure and the confirmation of crystal quality via the sharpness of Laue back-reflection spots are documented in [21]. Both samples were mechanically polished with successively finer abrasive paper (220 to 600 grit) and then chemically polished in an HF:$HNO_3$:glycerin (5:45:50) solution to remove the worked surface layer. Once mounted in the analysis chamber, residual carbon and oxygen contamination was removed by in-situ $Ar^+$ ion sputtering (5 keV, 4.5 keV focusing voltage, 10–20 μA sample current) until the C 1s and O 1s XPS signals were no longer detectable.

### 2.2. XPS measurements and oxidation procedure

Measurements were performed in a VG Scientific ESCA 3 Mk II electron spectrometer equipped with a hemispherical electrostatic analyser, a channeltron detector, twin Mg/Al X-ray anodes and two $Ar^+$ ion guns. The base pressure of the analysis and preparation chambers, each pumped by an oil-diffusion pump with a liquid-nitrogen trap, was $5 \times 10^{-10}$ torr. Spectra were excited with unmonochromated Mg Kα radiation (hν = 1253.6 eV) and binding energies were referenced to the Fermi level and calibrated against the Au 4f7/2 line at 84.0 eV.

The two crystals were mounted side by side and oxidized simultaneously, in the preparation chamber, at room temperature and at $O_2$ partial pressures of $1 \times 10^{-8}$, $1 \times 10^{-7}$ and $1 \times 10^{-6}$ torr (research-grade $O_2$, > 99.8% pure, with $H_2O$ < 5 ppm and $N_2$ + Ar < 2000 ppm). For each pressure, exposures were cumulative (no intermediate cleaning) and, after each exposure step, both samples were transferred to the analysis chamber, where narrow-scan Zr 3d and O 1s spectra were acquired (dwell time 1 s/channel for the kinetics measurements). Oxygen exposures are quoted in Langmuir units (1 L = $10^{-6}$ torr·s). To determine the in-depth distribution of the Zr–O compounds formed, additional Zr 3d and O 1s spectra were recorded, on the samples oxidized to saturation, as a function of the photoelectron take-off angle α (measured from the sample surface) between 20° and 60°; the counting time per channel was increased (3–5 s) at the lowest angles to compensate for the reduced signal. At the take-off angles used here (α = 20°–60°), this corresponds to an effective XPS sampling depth $\Lambda = \lambda \sin\alpha$ of approximately 6–16 Å; the non-angle-resolved measurements of Sections 3.1–3.2, acquired at a different, fixed detection geometry, sample to a depth of the order of λ itself (≈18 Å), comparable to, or somewhat greater than, the oxide-film thickness (10–19 Å), so that the underlying metal also contributes to the signal.

### 2.3. Data analysis

Atomic percentages of O and Zr were obtained from the Zr 3d and O 1s peak areas after Shirley background subtraction [22], using the Scofield photo-ionization cross-sections ($\sigma_{Zr} = 2.25$, $\sigma_O = 0.624$, relative to F 1s) [23]; the estimated uncertainty of this quantification, dominated by the signal-to-noise ratio of the spectra, is 2-3%. Because the Zr 3d spin–orbit doublet developed additional structure as the samples oxidized, the peak was deconvoluted using XPSPeak (R. W. M. Kwok, Department of Chemistry, The Chinese University of Hong Kong), which fits Gaussian–Lorentzian product line shapes by a least-squares procedure. The line shape, FWHM and doublet separation of the metallic-Zr and $ZrO_2$ components were fixed from reference spectra of a clean single crystal and of bulk $ZrO_2$, respectively; a satisfactory fit of the oxidized spectra required two additional doublets of intermediate binding energy, attributed to sub-oxide compounds and labelled (ZrO)a and (ZrO)b in order of increasing binding energy [4]. For the two sub-oxide components, both the binding energy and FWHM were left as free fit parameters; the narrow ranges quoted in Table 1 (±0.05 eV in binding energy, ±0.1 eV in FWHM) represent the spread of best-fit values obtained across all exposure steps, pressures and orientations studied. This four-component scheme, rather than the single continuously-shifting-binding-energy model proposed by other authors [2], follows the two-discrete-suboxide picture reported independently by [3,5] and by our own group for polycrystalline Zr [4]. The parameters used for all four components are given in Table 1. Oxide film thicknesses were calculated from the attenuation of the metallic-Zr 3d signal beneath the growing oxide layer using the expression of Kumar et al. [5], with an inelastic mean free path $\lambda = 18$ Å (escape-depth) and a metal-to-$ZrO_2$ density ratio $C = 0.63$.

| Component | Zr 3d5/2 B.E. (eV) | FWHM 3d5/2 (eV) | Spin–orbit splitting (eV) |
|---|---|---|---|
| Zr metal | 178.9 ± 0.1 | 1.3 (fixed) | 2.4 (fixed) |
| (ZrO)a | 180.2 ± 0.05 | 1.7 ± 0.1 | 2.4 ± 0.1 |
| (ZrO)b | 181.2 ± 0.05 | 1.7 ± 0.1 | 2.4 ± 0.1 |
| $ZrO_2$ | 182.7 ± 0.1 | 1.7 (fixed) | 2.4 (fixed) |

*Table 1. Binding energy (referenced to the Fermi level), FWHM and spin–orbit splitting used to fit the Zr 3d5/2,3/2 doublet of each chemical component.*

## 3. Results

### 3.1. Oxidation kinetics

At all three $O_2$ pressures and for both crystallographic orientations, the atomic percentage of oxygen at the surface increased with exposure in three well-defined stages: a fast initial uptake, a slower second stage, and a final saturation stage in which no further change was observed within the exposure range studied. This behavior is consistent with a logarithmic-type oxidation law, as expected for metal oxidation at low $O_2$ pressure and room temperature [1,13]. The duration of the first two stages decreased as the oxidation pressure increased, and at $1 \times 10^{-6}$ torr the first stage was too fast to be resolved with the present time resolution.

At every pressure, the near-prismatic sample (Z2) incorporated oxygen faster and reached a higher saturation concentration than the basal sample (Z1): approximately 42% at. O for Z1 versus 54% at. O for Z2 at $1\times10^{-8}$ torr; 50% versus 61% at $1\times10^{-7}$ torr; and 57% versus 65% at $1\times10^{-6}$ torr. In every case the difference between the two orientations was already present after the first exposure step, showing that the anisotropy in oxygen uptake sets in from the earliest stage of the interaction rather than developing only once a continuous oxide film has formed (Fig. 1). Because Z1 and Z2 were cut from the same parent single crystal and oxidized simultaneously, compositional and impurity differences between them are excluded by construction, isolating crystallographic orientation as the relevant variable. The same orientation-dependent ordering was reproduced independently at all three $O_2$ pressures studied, spanning two orders of magnitude, providing an internal consistency check.

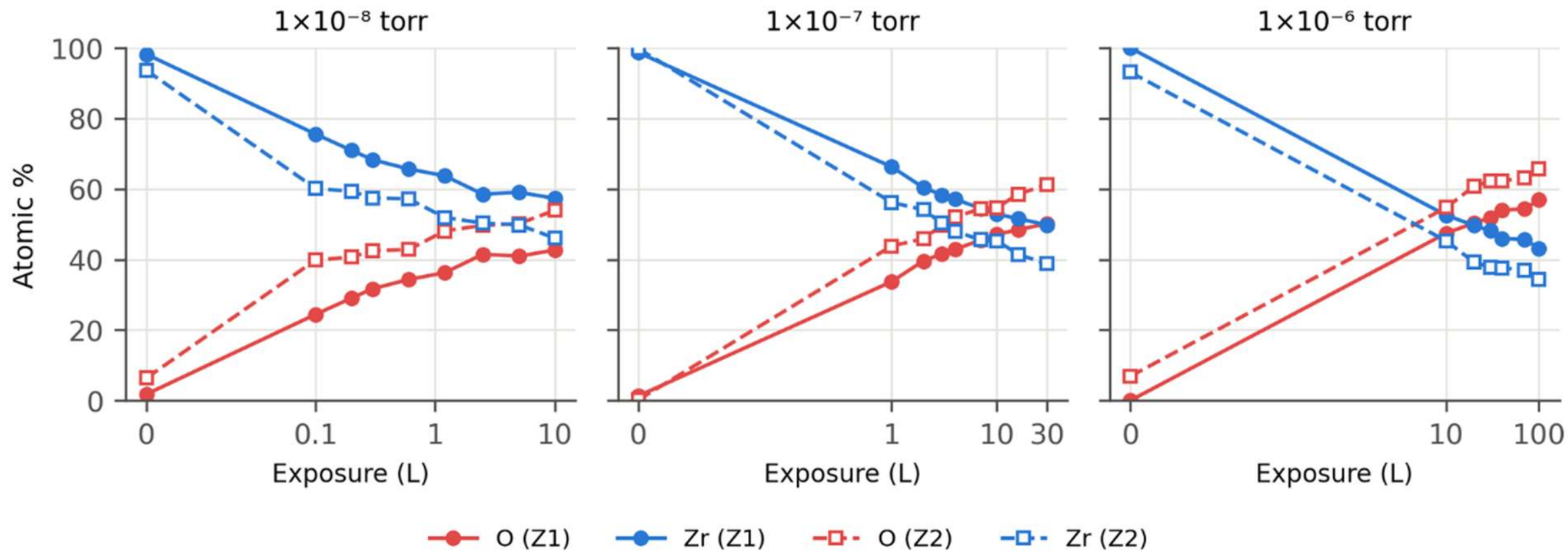

*Figure 1. Oxidation kinetics: O and Zr atomic percentage of the Zr 3d + O 1s signal vs. $O_2$ exposure, for Z1 (filled symbols) and Z2 (open symbols), at the three $O_2$ pressures studied.*

### 3.2. Chemical composition of the oxide films

As oxidation proceeded, the Zr 3d doublet progressively deformed towards higher binding energy (Fig. 2). A satisfactory fit of the experimental spectra at every exposure step, for both samples and all three pressures, required the four components listed in Table 1: metallic Zr (178.9 ± 0.1 eV), two sub-oxides (ZrO)a (180.2 ± 0.05 eV) and (ZrO)b (181.2 ± 0.05 eV) — shifted, respectively, by 1.3 and 2.3 eV from metallic Zr — and $ZrO_2$ (182.7 ± 0.1 eV). Both sub-oxides were already present after the very first exposure to $O_2$, well before the oxide film reached saturation, indicating that Zr–O bond formation begins immediately upon exposure rather than being preceded by a distinct, purely physisorbed or dissociatively chemisorbed stage without any charge transfer.

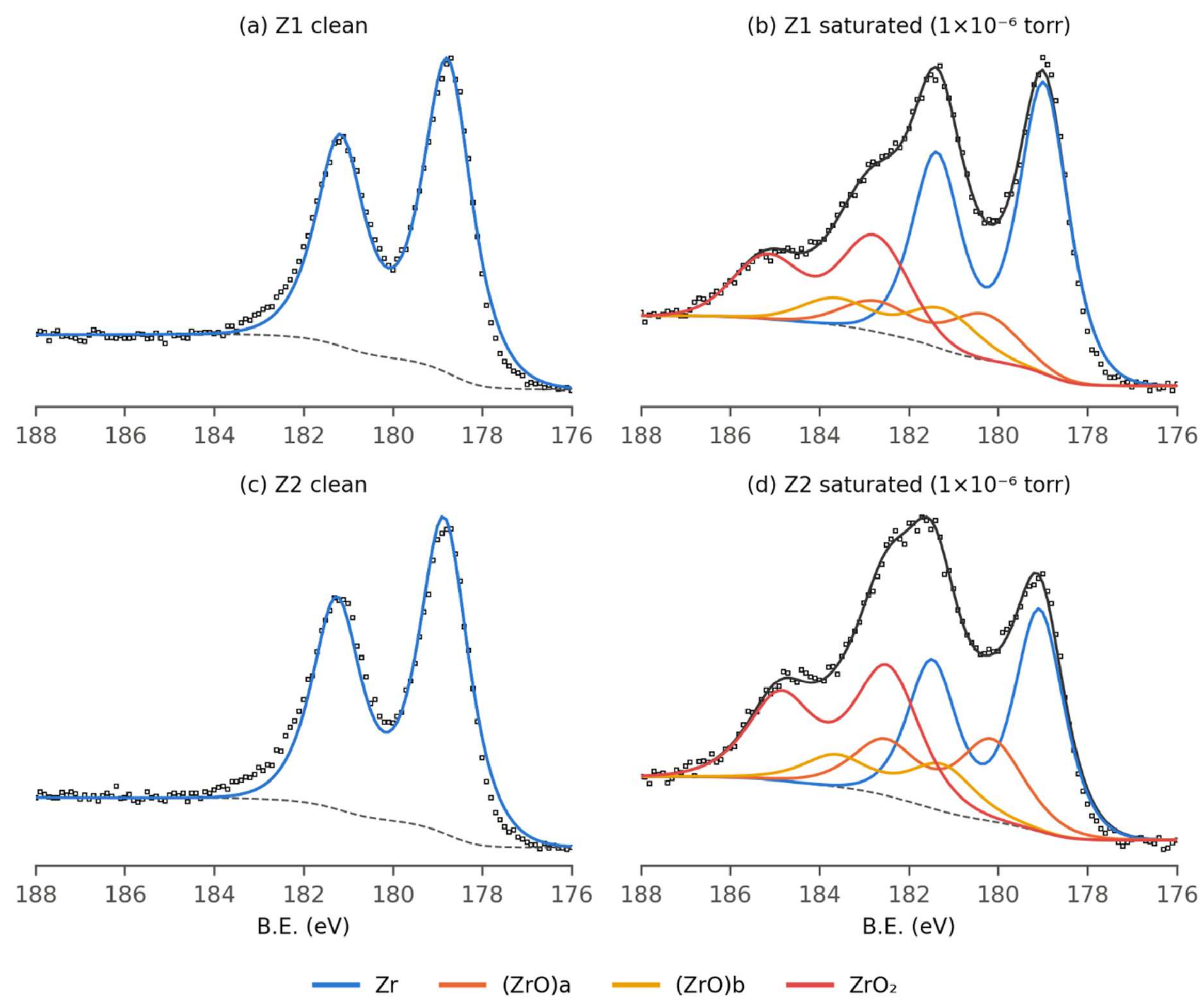

*Figure 2. Representative Zr 3d spectra and four-component deconvolution for the clean and saturated (1 × 10⁻⁶ torr) surfaces of Z1 and Z2.*

(ZrO)a was consistently the first compound to appear and the fastest to reach its limiting concentration, remaining thereafter essentially constant; (ZrO)b and $ZrO_2$ evolved more slowly. At saturation and $1 \times 10^{-8}$ torr, $ZrO_2$ slightly exceeds the suboxides concentration of the Z1 film but is comparable to, or exceeds the sub-oxides in the Z2 film; at $1 \times 10^{-7}$ torr, $ZrO_2$ became the largest oxidized component in Z1, while in Z2 it exceeded (ZrO)a only towards the end of the exposure; and at $1 \times 10^{-6}$ torr the amount of $ZrO_2$ was, for the first time, clearly the largest oxidized component in both orientations (Table 2; Fig. 3). Across all pressures and in both samples, (ZrO)a reached a similar saturation concentration in Z1 (≈ 12%) but, in Z2, saturated at roughly twice the concentration of (ZrO)b (≈ 19–22% vs. ≈ 9–12%); in Z1, by contrast, the two sub-oxides reached comparable final concentrations (≈ 10–12% each). In both orientations, metallic Zr remained the largest single component of the Zr 3d signal even at the highest oxidation pressure studied, confirming that the oxide film stayed thin enough for the substrate to remain detectable throughout.

| $O_2$ pressure | Sample | Zr (%) | (ZrO)a (%) | (ZrO)b (%) | $ZrO_2$ (%) |
|---|---|---|---|---|---|
| 1 × $10^{-8}$ torr | Z1 | 63.7 | 11.8 | 11.0 | 13.4 |
| | Z2 | 52.0 | 18.9 | 12.0 | 17.0 |
| 1 × $10^{-7}$ torr | Z1 | 59.6 | 12.0 | 10.4 | 17.9 |
| | Z2 | 43.8 | 21.5 | 8.6 | 26.0 |
| 1 × $10^{-6}$ torr | Z1 | 55.0 | 11.2 | 10.1 | 23.7 |
| | Z2 | 39.4 | 19.1 | 10.9 | 30.5 |

*Table 2. Atomic-percentage composition of the Zr 3d signal (metallic Zr and the three Zr–O components) at saturation, for both orientations and the three $O_2$ pressures studied, from the author's original peak-area quantification.*

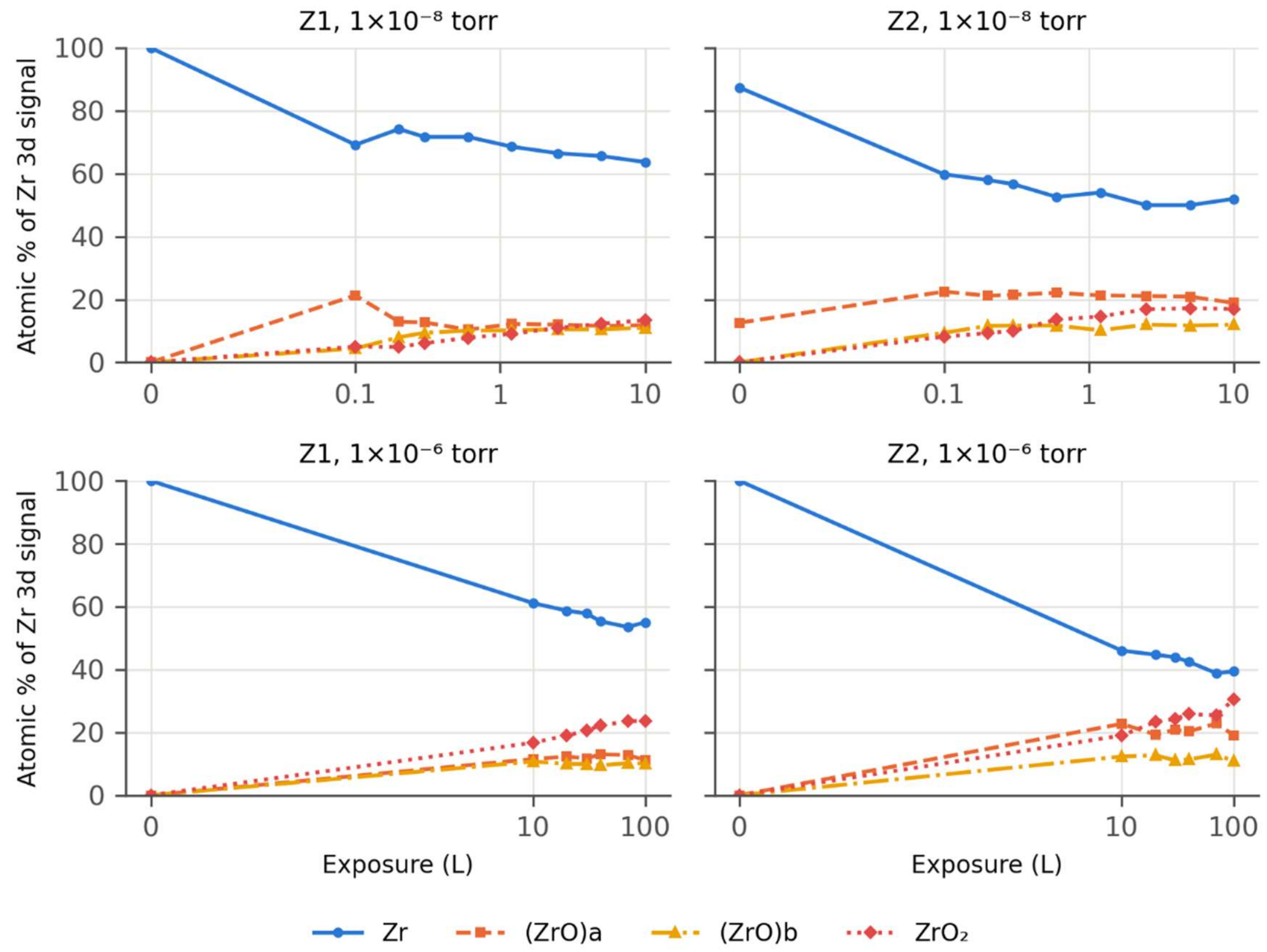

*Figure 3. Evolution of the four Zr 3d components (percentage of the total Zr 3d signal) with $O_2$ exposure, for Z1 and Z2 at 1 × $10^{-8}$ and 1 × $10^{-6}$ torr.*

## 3.3. In-depth distribution of the Zr–O compounds

The location of each Zr–O compound within the oxide film was investigated by recording Zr 3d spectra, on the saturated films, as a function of the photoelectron take-off angle α, which controls the effective sampling depth ($\Lambda = \lambda \sin\alpha$). In both orientations and at every pressure studied, the relative contribution of $ZrO_2$ to the Zr 3d signal increased markedly as α decreased (i.e. as the measurement became more surface-sensitive), while the metallic-Zr contribution decreased correspondingly (Fig. 4); this effect was more pronounced for Z1 than for Z2. The contributions of the two sub-oxides, in contrast, remained essentially constant as α was varied. These observations show that $ZrO_2$ is the outermost compound of the film in both orientations, consistent with the general picture of metal oxidation in which the most oxidized compound forms in contact with the gas phase, and with the less-oxidized ZrO sub-oxide phase reported closer to the metal–oxide interface in recent studies of corroded Zr alloys [17], while the two sub-oxides are distributed nearly homogeneously between the metal substrate and the outer $ZrO_2$ layer. This is consistent with a growth sequence in which oxygen incorporated at each new exposure first forms (ZrO)a, which is progressively converted into (ZrO)b and then into $ZrO_2$, with the flux of newly arriving oxygen continuously regenerating the reduced sub-oxide near the metal interface.

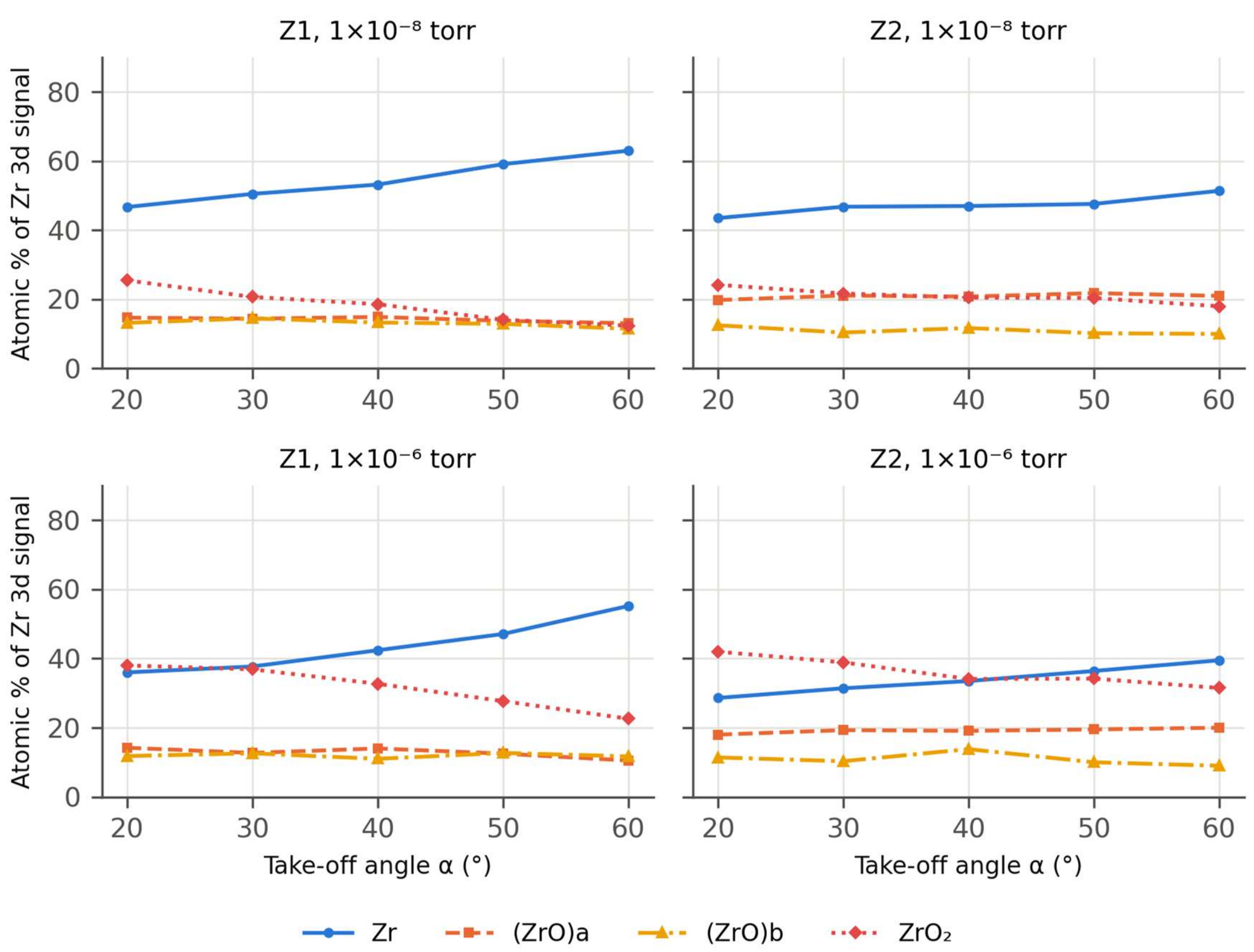

*Figure 4. Angle-resolved composition of the Zr 3d signal at saturation, as a function of the photoelectron take-off angle α (measured from the surface), for Z1 and Z2 at $1 \times 10^{-8}$ and $1 \times 10^{-6}$ torr.*

## 3.4. Oxide film thickness

Oxide thicknesses, calculated as described in Section 2.3, are given in Table 3 and Fig. 5. In both orientations the thickness increased rapidly with the first exposures and then levelled off to a pressure-dependent limiting value; this limiting thickness increased with oxidation pressure and, at every pressure, was larger for Z2 than for Z1. None of the films exceeded 20 Å, so that the metallic substrate remained within the XPS sampling depth throughout, consistent with the results of Section 3.2. The small non-zero thickness value obtained for Z2 at the very first exposure step at $1{\times}10^{-8}$ torr (Fig. 5) reflects the same faster initial oxygen incorporation on the near-prismatic orientation already noted in Section 3.1. The thicknesses obtained for Z2 agree closely with values previously reported for polycrystalline Zr oxidized under comparable conditions in the same laboratory [4] (Table 3); this can be attributed to the fact that cold-rolled

Zr tubes of the kind typically used in these earlier studies develop a strong crystallographic texture in which most grains present a prismatic, rather than basal, plane at the surface. This texture was described qualitatively in [4]. A quantitative, texture-resolved comparison of the kind shown by Perlovich et al. [24] to correlate with oxide growth rate, remains a valuable direction for future work.

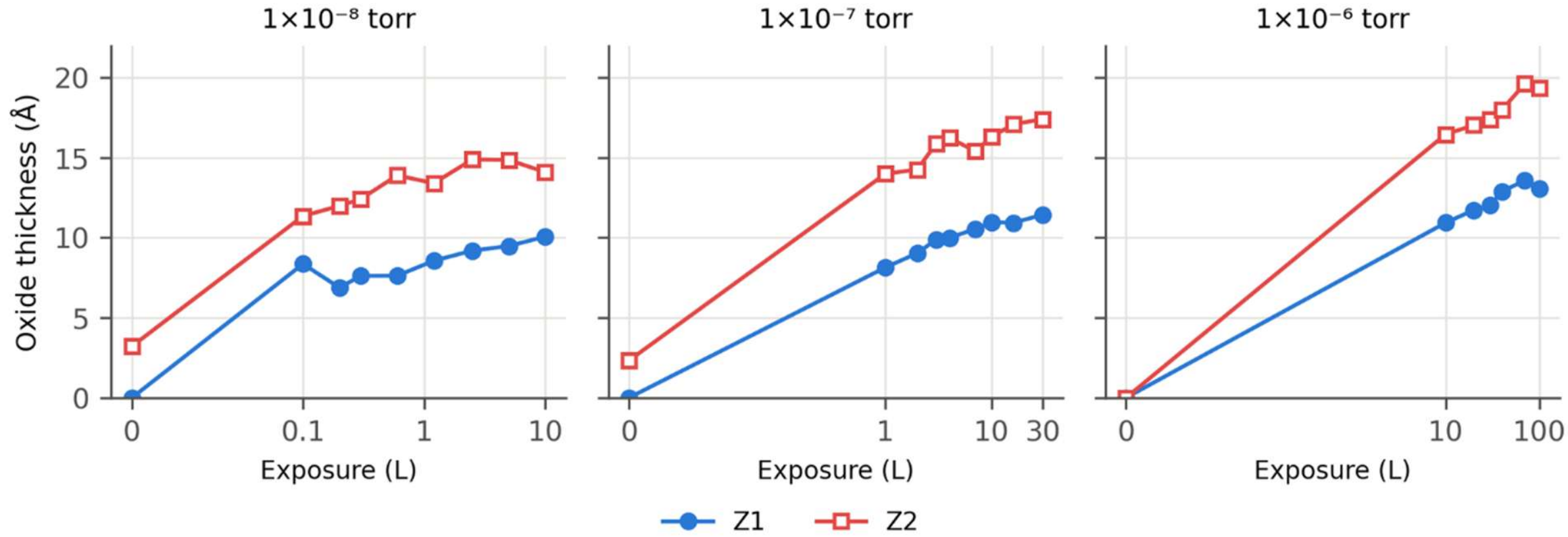

*Fig. 5. Oxide film thickness vs. $O_2$ exposure for Z1 and Z2 at the three $O_2$ pressures studied, from the attenuation of the metallic Zr 3d signal (Kumar expression [5], $\lambda = 18$ Å, $C = 0.63$).*

| $O_2$ pressure (torr) | Z1 (0001, basal) (Å) | Z2 (near-prismatic) (Å) | Polycrystalline Zr [4] (Å) |
|---|---|---|---|
| $1 \times 10^{-8}$ | 10.1 ± 2 | 14.1 ± 2 | 15 ± 2 |
| $1 \times 10^{-7}$ | 11.4 ± 2 | 17.4 ± 2 | — |
| $1 \times 10^{-6}$ | 13.1 ± 2 | 19.3 ± 2 | 19 ± 2 |

*Table 3. Oxide film thickness (Å) at saturation for Z1 and Z2 at the three $O_2$ pressures studied, compared with values reported for polycrystalline Zr under comparable conditions [4].*

## 4. Discussion

The three-stage, logarithmic oxidation kinetics observed here for both crystallographic orientations agree with the general picture established for polycrystalline Zr under similar conditions of low $O_2$ pressure and room temperature [3,4,13]. There is no full consensus in the literature, however, on the microscopic process underlying each stage: some authors [3,5] attribute the first stage to purely dissociative chemisorption of $O_2$ without Zr–O bond formation, followed by nucleation and then growth of oxide islands in the second and third stages, while others [4,5,7] report XPS evidence of Zr–O compound formation from the very first exposure. The present results agree with the latter picture: both sub-oxides were detected after the smallest exposure measured, at every pressure and in both orientations. This is consistent with the large chemical-potential gradient that transition metals such as Zr can sustain at their surface, which allows oxygen to migrate several atomic layers below the surface; recent atomic-resolution studies of the Zr-ZrO2 interface report structural features (broken interfacial coherency, Zr vacancies and nanopores) argued to facilitate this mass transport [16].

The binding energies and relative evolution of the two sub-oxide components identified here, (ZrO)a and (ZrO)b, match closely those previously reported by this group for polycrystalline Zr oxidized under comparable conditions [4]. Unlike that polycrystalline study, which reports orientation-averaged, saturated compositions, the present work isolates the orientation variable directly, on individually indexed single crystals, and follows the sub-oxide composition through many discrete exposure steps from the clean surface to saturation, complementing the paired Zr(0001)/Zr(10-10) kinetics study of Bakradze et al. [12] with this room-temperature, exposure-resolved account of how the sub-oxide composition itself evolves, on an orientation representative of the near-prismatic texture found in cold-worked cladding tubes. The chemical identity and even the number of Zr sub-oxide phases, however, remains a matter of debate: proposed assignments range from a single ZrOx compound with a continuously varying binding energy [2] to three discrete sub-oxide phases [3,10], and attempts to assign specific stoichiometries ($Zr_2O$, ZrO, $Zr_2O_3$) from the magnitude of the observed chemical shifts [3,5] rely on an empirical rule of thumb that, as noted by the authors of [5] themselves, has no general theoretical justification. This is not merely a historical disagreement: two independent studies published in the same 2022 volume of the Journal of Nuclear Materials debate the presence conditions and structural role of a ZrO sub-oxide phase at the metal–oxide interface of corroded Zr alloys [10,17,20], invoking hexagonal or cubic ZrO phases, point-defect aggregation, or metastable zirconia precursors to explain the observed deviation from $ZrO_2$ stoichiometry near the interface, and a 2026 review of Zr-alloy oxidation for nuclear cladding applications still describes the identity and structural role of sub-oxide phases at the metal–oxide interface as an open, and in places contradictory, question in the current literature [20]. The two well-resolved, angularly homogeneous sub-oxide components reported here, obtained on chemically and structurally well-defined single-crystal surfaces free of the grain-boundary and texture effects present in polycrystalline samples, remain a relevant data point for that discussion. We note that the present study, like most of the XPS literature on this system, characterizes the sub-oxide components solely by their electronic/chemical signature. The oxide films studied here are only 10–19 Å thick (Table 3), below the ~5 nm practical detection floor of even grazing-incidence XRD for well-ordered epitaxial films. Also, native oxides nucleating at room temperature without post-growth annealing are generally amorphous or, at best, nanocrystalline, with domain sizes likely too small to produce coherent Bragg scattering [25].

The systematic difference in oxidation extent, rate and film thickness between the near-prismatic (Z2) and basal (Z1) orientations observed here from the earliest exposures onward, is consistent with the more open atomic packing of the prismatic planes of the α-Zr hexagonal-close-packed structure relative to the more closely packed basal plane, which should offer a lower resistance to the inward migration of oxygen atoms and adsorbed species [3]. This packing argument is supported quantitatively by density-functional-theory calculations of oxygen migration into the basal and prism surfaces of Zr [14], which find a higher energy barrier for oxygen to penetrate the basal surface than the prism surface and a lower barrier for oxygen to escape from the basal surface than from the prism surface. Both effects are favoring net oxygen incorporation on the prism-type surface, which is consistent with the faster oxidation observed here for the near-prismatic orientation (Z2). The possible additional role of anisotropic oxygen self-diffusion along different crystallographic directions in α-Zr, and of dislocation structure, remains an open question for dedicated future studies combining precisely indexed, dislocation-characterized crystals with depth-resolved diffusion measurements.

More broadly, the extent to which crystallographic texture controls the macroscopic oxidation rate of Zr cladding alloys is, again, still described as unsettled in the most recent literature, with some studies reporting no measurable effect of texture on oxide microstructure and others reporting substantially different oxidation rates for differently textured material [20]. The close agreement found here between the oxide thickness of the near-prismatic single crystal (Z2) and that of cold-worked polycrystalline Zr tubes, whose surface is dominated by prismatic-oriented grains [4], is a further indication that crystallographic orientation is a relevant, first-order variable in the early-stage oxidation of α-Zr and should be considered explicitly when comparing oxidation data obtained on differently textured material.

## 5. Conclusions

The early-stage, room-temperature oxidation of two α-Zr single crystals of different, well-characterized crystallographic orientation — basal (Z1) and near-prismatic (Z2) — was studied by XPS at $O_2$ pressures between $1 \times 10^{-8}$ and $1 \times 10^{-6}$ torr. In both orientations the oxidation kinetics followed a three-stage, logarithmic-type law, but the near-prismatic sample consistently incorporated oxygen faster and to a greater final extent than the basal sample, at every pressure studied and from the earliest exposures. Deconvolution of the Zr 3d spectra resolved two sub-stoichiometric Zr–O compounds, in addition to metallic Zr and $ZrO_2$, with binding energies and relative evolution consistent with sub-oxides previously reported for polycrystalline Zr, confirming their presence on well-defined single-crystal surfaces. Angle-resolved XPS showed that $ZrO_2$ is the outermost compound of the oxide film in both orientations, while the two sub-oxides are distributed nearly homogeneously through the film thickness. The limiting oxide thickness, between 10 and 19 Å depending on orientation and pressure, was systematically larger for the near-prismatic than for the basal orientation, and closely matched literature values for cold-worked polycrystalline Zr, whose surface texture favors prismatic orientations. Taken together, these results establish that crystallographic orientation is a determining factor in the early-stage oxidation of Zr, governing the oxidation kinetics, the sub-oxide-to-$ZrO_2$ balance, and the limiting oxide thickness alike and, by directly comparing two well-defined orientations of the same parent single crystal, provide clear evidence that surface texture is central to understanding the anisotropic oxidation behavior of Zr and its alloys in nuclear applications.

## Data availability

The data that support the findings of this study are available from the corresponding author upon reasonable request.

## Acknowledgements

This work was funded by ANPCyT (PICT-2021-495) and CONICET (PIP-0516/2022).

The authors gratefully acknowledge the sustained state policies that, for decades, supported scientific and technological development in Argentina and at CNEA in particular, and made this and much other work possible. We note with concern that these policies are currently being eroded by a substantial reduction in resources — both human and financial — in a field, nuclear science and technology, that should be treated as strategic. We believe this is a matter of legitimate concern for the broader scientific community, and remain open to engaging with colleagues and institutions working to address it, in whatever way may be useful.

## Declaration of Generative AI and AI-assisted technologies in the writing process

During the preparation of this work, the authors used an AI-based language assistant (Claude, Anthropic) in order to improve the clarity and English-language phrasing of the manuscript text and of the response to reviewers. After using this tool, the authors reviewed and edited the content as needed and take full responsibility for the content of the publication.

## References

[1] R.L. Tapping, X-ray photoelectron and ultraviolet photoelectron studies of the oxidation and hydriding of zirconium, J. Nucl. Mater. 107 (1982) 151–158. https://doi.org/10.1016/0022-3115(82)90417-2.

[2] P. Sen, D.D. Sarma, R.C. Budhani, K.L. Chopra, C.N.R. Rao, An electron spectroscopic study of the surface oxidation of glassy and crystalline Cu-Zr alloys, J. Phys. F: Met. Phys. 14 (1984) 565–577. https://doi.org/10.1088/0305-4608/14/2/027.

[3] J.M. Sanz, C. Palacio, Y. Casas, J.M. Martínez-Duart, An AES study of the oxidation of polycrystalline zirconium at room temperature and low oxygen pressures, Surf. Interface Anal. 10 (1987) 177. https://doi.org/10.1002/sia.740100402.

[4] C. Oviedo de González, E.A. García, An X-ray photoelectron spectroscopy study of the surface oxidation of zirconium, Surf. Sci. 193 (1988) 305–320. https://doi.org/10.1016/0039-6028(88)90438-4.

[5] L. Kumar, D.D. Sarma, S. Krummacher, XPS study of the room temperature surface oxidation of zirconium and its binary alloys with tin, chromium and iron, Appl. Surf. Sci. 32 (1988) 309–319. https://doi.org/10.1016/0169-4332(88)90016-5.

[6] C.S. Zhang, B. Flinn, I.V. Mitchell, P.R. Norton, The initial oxidation of Zr(0001): 0 to 0.5 monolayers, Surf. Sci. 245 (1991) 373–379. https://doi.org/10.1016/0039-6028(91)90039-U.

[7] G. Bakradze, L.P.H. Jeurgens, E.J. Mittemeijer, Valence-band and chemical-state analyses of Zr and O in thermally grown thin zirconium-oxide films: an XPS study, J. Phys. Chem. C 115 (2011) 19841–19848. https://doi.org/10.1021/jp206896m.

[8] Z. Azdad, L. Marot, L. Moser, R. Steiner, E. Meyer, Valence band behaviour of zirconium oxide, photoelectron and Auger spectroscopy study, Sci. Rep. 8 (2018) 16251. https://doi.org/10.1038/s41598-018-34570-w.

[9] H. Li, J.-I.J. Choi, W. Mayr-Schmölzer, C. Weilach, C. Rameshan, F. Mittendorfer, J. Redinger, M. Schmid, G. Rupprechter, Growth of an ultrathin zirconia film on $Pt_3Zr$ examined by high-resolution X-ray photoelectron spectroscopy, temperature-programmed desorption, scanning tunneling microscopy, and density functional theory, J. Phys. Chem. C 119 (2015) 2462–2470. https://doi.org/10.1021/jp5100846.

[10] L. Chen, B. Luan, S. Ma, P. Wan, G. Bai, Y. Liu, Y. Zhang, Discussion on the presence condition of suboxide ZrO beneath the oxide in zirconium alloys, J. Nucl. Mater. 571 (2022) 154011. https://doi.org/10.1016/j.jnucmat.2022.154011.

[11] J.P. Pemsler, Diffusion of Oxygen in Zirconium and its Relation to Oxidation and Corrosion, J. Electrochem. Soc. 105 (1958) 315–322. https://doi.org/10.1149/1.2428837

[12] G. Bakradze, L.P.H. Jeurgens, E.J. Mittemeijer, The different initial oxidation kinetics of Zr(0001) and Zr(10-10) surfaces, J. Appl. Phys. 110 (2011) 024904. https://doi.org/10.1063/1.3608044

[13] Y. Tang, J. Liao, D. Yun, Understanding the high-temperature corrosion behavior of zirconium alloy as cladding tubes: a review, Front. Mater. 11 (2024) 1381818. https://doi.org/10.3389/fmats.2024.1381818.

[14] T.-W. Chiang, A. Chernatynskiy, M.J. Noordhoek, S.B. Sinnott, S.R. Phillpot, Anisotropy in oxidation of zirconium surfaces from density functional theory calculations, Comput. Mater. Sci. 98 (2015) 112–116. https://doi.org/10.1016/j.commatsci.2014.10.052.

[15] E.J. Kautz, B. Gwalani, S.V.M. Lambeets, L. Kovarik, D.K. Schreiber, D.E. Perea, D. Senor, Y.-S. Liu, A.K. Battu, K.-P. Tseng, S. Thevuthasan, A. Devaraj, Rapid assessment of structural and compositional changes during early stages of zirconium alloy oxidation, npj Mater. Degrad. 4 (2020) 29. https://doi.org/10.1038/s41529-020-00133-6.

[16] Y. Zhang, Z. Li, X. Tong, Z. Xie, S. Huang, Y.-E. Zhang, H.-B. Ke, W.-H. Wang, J. Zhou, Three-dimensional atomic insights into the metal-oxide interface in Zr-$ZrO_2$ nanoparticles, Nat. Commun. 15 (2024) 7624. https://doi.org/10.1038/s41467-024-52026-w.

[17] T. Wei, X. Dai, Y. Zhao, D. Wang, J. Lv, Y. Huang, J. Zhang, ZrO phase embedded in the oxide of Zr-Sn-Nb-Fe-Cr alloy after corrosion, J. Nucl. Mater. 571 (2022) 153992. https://doi.org/10.1016/j.jnucmat.2022.153992.

[18] I. Betova, M. Bojinov, V. Karastoyanov, Long-term oxidation of zirconium alloy in simulated nuclear reactor primary coolant—experiments and modeling, Materials 16 (2023) 2577. https://doi.org/10.3390/ma16072577.

[19] J. Hartmann, T. Varga, C. Schenck, C. McRobie, F.-Y. Tsai, V. Shutthanandan, A. Devaraj, D. Senor, B. Gwalani, E. Kautz, Structure evolution and tin redistribution during oxidation of Zircaloy-4 at 500 °C, J. Nucl. Mater. 614 (2025) 155895. https://doi.org/10.1016/j.jnucmat.2025.155895.

[20] T.-Y. Liu, W.-Z. Han, Oxidation of zirconium alloys for nuclear fuel cladding, Commun. Mater. 7 (2026) 137. https://doi.org/10.1038/s43246-026-01201-1.

[21] S.N. Balart, N. Varela, R.H. de Tendler. $^{51}$Cr DIFFUSION IN α-Zr SINGLE CRYSTALS. Journal of Nuclear Materials 119 (1983) 59-66. https://doi.org/10.1016/0022-3115(83)90052-1

[22] D.A. Shirley, High-resolution X-ray photoemission spectrum of the valence bands of gold, Phys. Rev. B 5 (1972) 4709–4714. https://doi.org/10.1103/PhysRevB.5.4709.

[23] J.H. Scofield, Hartree-Slater subshell photoionization cross-sections at 1254 and 1487 eV, J. Electron Spectrosc. Relat. Phenom. 8 (1976) 129–137. https://doi.org/10.1016/0368-2048(76)80015-1.

[24] Yu.A. Perlovich, M.G. Isaenkova, P.N. Medvedev, V.A. Fesenko, Soe San Thu, Mechanisms of the texture influence on the corrosion behavior of Zr-alloy cladding tubes, Inorg. Mater. Appl. Res. 6 (2015) 259–266. https://doi.org/10.1134/S2075113315030077.

[25] D.L. Douglass, The Metallurgy of Zirconium, Atomic Energy Review, Supplement 1971, International Atomic Energy Agency, Vienna, 1971. IAEA sales no. STI/PUB/066/S/1971, ISBN 92-0-159071-7.